\documentclass[aps,pra,showpacs,twocolumn,amsmath,amssymb,superscriptaddress,footinbib]{revtex4}

\usepackage[english]{babel}
\usepackage{latexsym}
\usepackage{graphics}
\usepackage{subfigure}
\usepackage{epsfig}
\usepackage{amsfonts}
\usepackage{amssymb}
\usepackage{amsmath}
\usepackage{bbm}

\begin{document}

\title{Analog of the spin-orbit induced anomalous Hall effect with quantized radiation}


\author{Jonas Larson}
\email{jolarson@kth.se} \affiliation{NORDITA, 106 91 Stockholm, Sweden}
\affiliation{Department of Physics, Stockholm University, AlbaNova University Center, 106 91 Stockholm, Sweden}
\date{\today}

\begin{abstract}
We demonstrate how the term describing interaction between a single
two-level atom and two cavity field modes may attain a Rashba form.
As an outcome, cavity QED provides a testbed for studies of phenomena reminiscent of the spin-orbit induced
anomalous Hall effect. The effective magnetic field, deriving from
the non-Abelian gauge potentials rendered by the Rashba coupling,
induces a transverse force acting on the phase space distributions.
Thereby, the phase space distributions build up a transverse motion manifesting itself in spiral trajectories,
rather than circular ones obtained for a zero magnetic field as one
would acquire for the corresponding Abelian gauge potentials.
Utilizing realistic experimental parameters, the phenomenon is
numerically verified, ascertain that it should be realizable with
current techniques.
\end{abstract}


\maketitle

Progress in manufacturing and a deeper understanding of electronic
devices have caused an increased interest in the field of
spintronics \cite{spintronics}. For example, the spin-orbit coupling
occurring in various materials such as semiconductors provides a
handle for the electron spin. In fact, the spin-orbit coupling
is a main ingredient for both the spin Hall effect
\cite{spinhall} and its companion, the anomalous Hall effect (AHE)
\cite{anhall}. This makes these effects attractive from a practical
point of view, and not only as phenomena on their own right.
However, despite its long history, the underlying physics of
especially the latter, the AHE, has remained a controversial subject
\cite{controverse}.

Already E. Hall discovered that the Hall effect is greatly enhanced
in ferromagnetic compared to non-magnetic materials \cite{ehall}.
This became known as the AHE. It is only recently, with knowledge of
geometrical phases, that a more complete picture of the AHE has been
obtained \cite{anhall}. The processes contributing to the AHE can be
divided into three categories: (i) intrinsic (deriving from the band
structure, and consequently from the spin-orbit coupling), (ii) skew
scattering (originating from distorted scattering of impurities),
and (iii) side-jump (often identified as the contribution making up
for the difference between the total AHE conductivity and the
intrinsic plus skew conductivities). While the first has a
topological origin, the last two contributions are due to impurities
in the material. Numerous measurements of the AHE conductivity have
been presented \cite{hallexp}, but identifying, and even
controlling, the different mechanisms underlying the AHE current
remains an extremely non-trivial task. In the realm of spintronics,
it would therefore be desirable to come up with alternative systems
free from impurities such that the AHE current derives solely from
the spin-orbit coupling, which could, for example, lead to accomplishing effective Datta-Das transistors~\cite{ddt}. Overcoming the obstacle of impurities with
solid state devices seems very difficult, and one should preferably
look in other directions.

Two possible candidates, both known to exhibit Hall characteristics,
are either rotating Bose-Einstein condensates \cite{BECHall} or cold
atoms in optical lattices \cite{OptlatHall}. Due to the purity and
high controlled versatility of these systems, great interest has
been paid to them during the recent past \cite{maciek}. The AHE has
been predicted to occur on the $p$-band of cold fermions in an
optical lattice \cite{pband}. An optional system considers cold
atoms coupled to spatially varying light fields \cite{gaugelaser},
in which the spin Hall effect has been predicted \cite{spinHallCA}.
Regardless of the development of novel experimental techniques in
cooling and trapping cold atom gases, presently non of the different
Hall effects nor non-Abelian properties have been realized utilizing cold atoms.

Yet another system endowed from impurities and possessing long
coherence time scales is cavity quantum electrodynamics (QED); a
single atom interacting with an isolated set of cavity modes
\cite{cavityqed}. During the last two decades, cavity QED has
successfully demonstrated among others, entanglement generation
\cite{cavityent}, the quantum measurement problem and the
quantum-classical transition \cite{cavitymeas}, and verification of
the graininess of the quantized electromagnetic field \cite{rabi}.
Moreover, recent experiments have realized coherent coupling of
single quantum dots \cite{cavitydot} or Bose-Einstein condensates
\cite{cavityBEC} to a cavity mode, paving the way for the
possibility of reaching a super-strong coupling regime of cavity
QED.

Notwithstanding the many advantages provided by cavity QED models,
relatively few schemes employing them as quantum simulators have
been suggested. In Ref.~\cite{jonas} it was presented how effective
non-Abelian gauge potentials arise in bimodal cavity QED models.
Similar cavity QED systems had also been shown to mimic Jahn-Teller
models frequently appearing in molecular and condensed matter
physics \cite{jonas2}. A proposal how to realize magnetic monopoles by the means of cavity QED systems was outlined in~\cite{cavitymonopole}. A model similar in structure, the cold
trapped ion system, was analyzed in~\cite{ion} and it was put
forward how it may simulate relativistic particles (see also
\cite{ionrev}). In this paper we demonstrate the appearance of transverse phase space currents, evocative of intrinsic
AHEs, in a bimodal cavity QED system.

We consider a high-$Q$ cavity containing a two-level atom
(quantum-dot) dipole-interacting with two degenerate but orthogonal
cavity modes. The system Hamiltonian reads
\begin{equation}\label{ham1}
\hat{H}=\hat{H}_f+\hat{H}_a+\hat{H}_I
\end{equation}
where
\begin{equation}
\begin{array}{ll}
\hat{H}_f=\displaystyle{\hbar\omega\left(\hat{a}^\dagger\hat{a}+\hat{b}^\dagger\hat{b}\right)}, & 
\hat{H}_a=\displaystyle{\sum_{j=1,2}E_j|j\rangle\langle j|},\\
\hat{H}_I=\displaystyle{\bar{d}\cdot\bar{E}(\mathbf{x})}. &
\end{array}
\end{equation}
Here, $\hat{a}^\dagger$ and $\hat{b}^\dagger$ ($\hat{a}$ and
$\hat{b}$) represent the creation (annihilation) operators for the
two modes, $\omega$ is their common mode frequency, $E_j$ the
internal atomic energy of state $|j\rangle$, $\bar{d}$ is the dipole
transition moment, and $\bar{E}({\bf x})$ is the electric field. In
terms of the internal atomic states, the components of the dipole
moment operator becomes
$d_\alpha=-e|1\rangle\alpha\langle2|-e|2\rangle\alpha\langle1|$,
where $e$ is the electron charge and $\alpha=x,y,z$. In the dipole
approximation we set $\mathbf{x}=0$, and the field is written as
\begin{equation}
\bar{E}=\bar{\varepsilon}_1\mathcal{E}_1i\left(\hat{a}-\hat{a}^\dagger\right)+\bar{\varepsilon}_2\mathcal{E}_2i\left(\hat{b}-\hat{b}^\dagger\right),
\end{equation}
where $\bar{\varepsilon}_j$ is the polarization vector for mode $j$
and $\mathcal{E}_j$ its corresponding field amplitude. A deeper
insight of the system characteristics is obtained by expressing the
fields in their quadrature operators
\begin{equation}\label{quad}
\begin{array}{l}
\hat{X}_1=\displaystyle{\frac{1}{\sqrt{2}}\left(\hat{a}+\hat{a}^\dagger\right)},\hspace{0.8cm}
\hat{P}_1=\displaystyle{\frac{i}{\sqrt{2}}\left(\hat{a}-\hat{a}^\dagger\right)},\\ 
\hat{X}_2=\displaystyle{\frac{1}{\sqrt{2}}\left(\hat{b}+\hat{b}^\dagger\right)},\hspace{0.8cm}
\hat{P}_2=\displaystyle{\frac{i}{\sqrt{2}}\left(\hat{b}-\hat{b}^\dagger\right)}
\end{array}
\end{equation}
obeying the canonical commutation relations;
$[\hat{X}_k,\hat{P}_{k'}]=i\delta_{kk'}$. For brevity we introduce
the Pauli operators
$\hat{\sigma}_x=|1\rangle\langle2|+|2\rangle\langle1|$,
$\hat{\sigma}_y=-i|1\rangle\langle2|+i|2\rangle\langle1|$,
$\hat{\sigma}_z=|1\rangle\langle1|-|2\rangle\langle2|$, and set the
zero energy such that $E_1=\hbar\Omega/2$ and $E_2=-\hbar\Omega/2$.
Furthermore, we choose
$\hat{g}_1\equiv\bar{d}\cdot\bar{\varepsilon}_1\mathcal{E}_1\sqrt{2}=-g\hat{\sigma}_x$,
and assume
$\hat{g}_2\equiv\bar{d}\cdot\bar{\varepsilon}_2\mathcal{E}_2\sqrt{2}=-g\hat{\sigma}_y$
with $g$ real. The resulting Hamiltonian is the $E\times\varepsilon$
Jahn-Teller one \cite{jonas2}
\begin{equation}\label{hamBR}
\begin{array}{lll}
\hat{H}_{E\times\varepsilon} & = &
\displaystyle{\hbar\omega\sum_{k=1,2}\left(\frac{\hat{P}_k^2}{2}+\frac{\hat{X}_k^2}{2}\right)+\frac{\hbar\Omega}{2}\hat{\sigma}_z}\\
& & -\displaystyle{\hbar
g\left(\hat{\sigma}_x\hat{P_1}+\hat{\sigma}_y\hat{P}_2\right)},
\end{array}
\end{equation}
first introduced in the molecular physics community~\cite{conical}. Via a simple
$\pi$-rotation around the $\hat{\sigma}_x$ axis, the atom-field
interaction term attains a Rashba form \cite{rashba}. In the absence of $\hat{X}_{1,2}$-terms, the same type of Hamiltonian is frequently used to describe a two dimensional gas of free spin-orbit coupled electrons. Adding the parabolic potential renders, on the other hand, a stereotype model representing spin-orbit coupled quantum dots~\cite{qdots}. Such confinement naturally restricts the particle motion, but nonetheless, as will be shown spin-orbit induced transverse currents still play a crucial role for the system dynamics. 

\begin{figure}[h]
\centerline{\includegraphics[width=4.2cm]{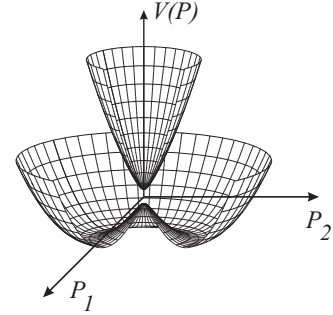}}
\caption{The APSs of the $E\times\varepsilon$ Hamiltonian
(\ref{hamBR}). The conical intersection is located
at the origin, $P_1=P_2=0$.} \label{fig1}
\end{figure}

The adiabatic potential surfaces (APS) of
$\hat{H}_{E\times\varepsilon}$, defined as
$V_{ad}^\pm(P_1,P_2)=\hbar\omega\left(P_1^2+P_2^2\right)/2\pm\hbar\sqrt{\Omega^2/4+g^2\left(P_1^2+P_2^2\right)}$,
are envisaged in Fig.~\ref{fig1}. These have a polar symmetry and
possess a conical intersection at the origin \cite{conical}. The $\hat{\sigma}_z$-term split
the degeneracy at the origin, and whenever $g<\sqrt{\omega\Omega}$
the sombrero shape of the lower APS is lost and instead a global
minimum at the origin is attained. Conical intersections appearing in momentum space are frequently referred to as Dirac points due to their linear dispersions.  

The Hamiltonian is conveniently written on the form
\begin{equation}\label{hamBR2}
\hat{H}_{E\times\varepsilon}=
\displaystyle{\hbar\omega\sum_{j=1,2}\left(\frac{\left(\hat{P}_j-\hat{A}_j\right)^2}{2}+\frac{\hat{X}_k^2}{2}\right)+\frac{\hbar\Omega}{2}\hat{\sigma}_z}+\hat{\Phi}
\end{equation}
where
\begin{equation}\label{vecpot}
(\hat{A}_1,\hat{A}_2)=\frac{g}{\omega}(\hat{\sigma}_x,\hat{\sigma}_y),\hspace{1cm}\hat{\Phi}=-\hbar\frac{g^2}{\omega}.
\end{equation}
The quantities $\hat{A}_k$ and $\hat{\Phi}$ are effective vector and
scalar potentials respectively. The gauge invariance follows
directly from considering their response to unitary transformations;
\begin{equation}\label{gaugetrans}
\begin{array}{l}
\hat{A}\rightarrow
\displaystyle{U^\dagger(\hat{X},t)\hat{A}U(\hat{X},t)-U^\dagger(\hat{X},t)\frac{\partial}{\partial
\hat{X}}U(\hat{X},t)},\\ 
\hat{\Phi}\rightarrow
\displaystyle{U^\dagger(\hat{X},t)\hat{\Phi} U(\hat{X},t)-i\hbar
U^\dagger(\hat{X},t)\frac{\partial}{\partial t}U(\hat{X},t)}.
\end{array}
\end{equation}
Pointed out in Ref.~\cite{berrycon}, the intrinsic AHE is a direct
result of a non-zero Berry phase originating from a vector potential
(also called Mead-Berry curvature). One way to illustrate the appearance
of the Hall effect is by the generalized spin-dependent Lorenz force analogous to the work by Wong on color-charged classical particles moving in non-Abelian fields~\cite{wong}.
As an outcome of the gauge potentials, there is an associating
magnetic field
\begin{equation}\label{magfield}
\hat{B}_i=\frac{1}{2}\varepsilon_{ijk}\hat{F}_{kl},\hspace{0.6cm}\hat{F}_{kl}=\partial_k\hat{A}_l-\partial_l\hat{A}_k-i[\hat{A}_k,\hat{A}_l].
\end{equation}
The first two terms of $\hat{F}_{kl}$ are identically zero in our
model. On the other hand, the last term, deriving from the
non-Abelian structure of the system \cite{wong}, is indeed non-zero. The
effective magnetic field, being proportional to $\hat{\sigma}_z$,
induces a state (spin) dependent force; the magnetic field is either
in the positive or negative $z$-direction depending on the internal
atomic state $|1\rangle$ or $|2\rangle$. Due to the opposite
direction of the force on the internal states, the system may act as
a spin-filter (Datta-Das transistor) which explains its interest for spintronics. Here, however, the intrinsic transverse AHE current does not take place in real space as
in solid state devices, but in a phase space representation of the
field states. 

For typical atom cavity QED parameters it is legitimate to apply the
rotating wave approximation, which would cast the Hamiltonian
(\ref{hamBR}) into a bimodal Jaynes-Cummings one \cite{jc}. On the
other hand, for circuit cavity QED employing solid state quantum
dots/SQUIDs instead of true atoms, imposing the rotating wave approximation
is usually not justified. In fact, in the parameter regimes where this approximation is applicable one finds that the transverse current rendered by the spin-orbit coupling is rather weak, and spontaneous emission of the atom as well as cavity decay will thereby become significant within the time needed to build up a noticeable transverse current.  

\begin{figure}[h]
\centerline{\includegraphics[width=7.0cm]{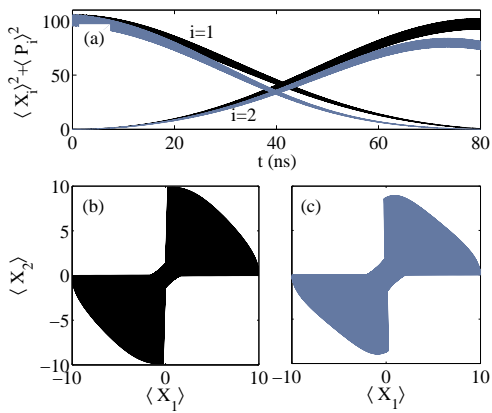}} \caption{(Color
online) The upper plot (a) displays the average radius of the phase
space distributions as function of time. Black curves shows the
ideal situation without losses, while for the light blue curve
losses have been taken into account. The two lower plots present the
same time evolution but of $\langle\hat{X}_1\rangle$ vs.
$\langle\hat{X}_2\rangle$ without (b) and with (c) losses. The
system parameters are $P_{10}=0$, $P_{20}=0$, $X_{10}=10$,
$X_{20}=0$, $\Omega/2\pi=6.9$ GHz, $\omega/2\pi=5.7$ GHz,
$g/2\pi=105$ MHz, $\gamma/2\pi=1.9$ MHz, and $\kappa/2\pi=250$ kHz.}
\label{fig2}
\end{figure}

Normally, the transverse AHE current is driven by an
electric field having the effect of tilting the 2D $X_1X_2$-plane.
In this cavity QED setting, a comparable tilt is generated by external
driving of the cavity. Pumping with an amplitude $\eta$ of say mode
1, implies a term
$\hat{H}_{pump}=\hbar\eta\left(\hat{a}_1+\hat{a}_1^\dagger\right)=\hbar\sqrt{2}\eta\hat{X}_1$
in the Hamiltonian (\ref{hamBR2}). This induces a position shift of
oscillator 1. An alternative way of picturing the pumping is to
displace the initial field state of mode 1 by the operator
$\hat{D}(\alpha)=e^{\alpha\hat{a}_1^\dagger-\alpha^*\hat{a}_1}$,
with $\alpha=\eta\sqrt{2}/\omega$. In our numerical simulations we
will consider this alternative viewpoint by assuming the initial
state of mode 1 to be displace by $\alpha$, whereby we omit any
pumping term $\hat{H}_{pump}$ in our numerical calculations.

In a solid, impurities prevents the electrons to accelerate to infinity causing an equilibrium situation. It is known that a constantly accelerating situation results in no net transverse current~\cite{jonasbloch}. The present model lacks impurities, but on the other hand the confining potential of Eq.~(\ref{hamBR2}) restricts the motion in the
$X_1X_2$-plane and the spin-orbit induced AHE is thereby still apparent in the phase space motion. 

To demonstrate the parallel of an intrinsic AHE in the present system we solve the Schr\"odinger equation
employing the split-operator wave packet method. In particular we
utilize the experimental parameters of the circuit QED
experiment presented in Ref.~\cite{cavitydot}, and moreover assume
the atom (quantum-dot) to be initially in its ground state
$|1\rangle$, mode 2 in vacuum $|0\rangle_2$, and mode 1 in a
coherent state $|\alpha_{10}\rangle_1$ (displaced vacuum);
\begin{equation}
\displaystyle{\psi_1(X_1,0)=\left(\frac{1}{\pi}\right)^{1/4}e^{-iP_{10}X_1}e^{-\left(X_1-X_{10}\right)^2/2}}\\ \\
\end{equation}
Here, $X_{10}$ and $P_{10}$ are respectively the initial position
and momentum of the coherent state, related to the amplitude as
$\alpha_{10}=(X_{10}-iP_{10})/\sqrt{2}$. By choosing $P_{10}=0$ and
$X_{10}>0$, the combined field wave packet starts out with a zero
momentum at $(X_1,X_2)=(X_{10},0)$. Apart from the potential force
causing the wave packet to slide down the harmonic potential, in
addition it experiences the effective transverse Lorentz force
originating from the magnetic field (\ref{magfield}). For an Abelian
gauge potential ($\hat{g}_1\propto\hat{g}_2$), the Lorentz force
vanishes and the field wave packet would remain localized along the
$X_1$-axis throughout the propagation. In the present model,
however, the oscillatory motion will depart from the $X_1$-axis
inducing a non-zero field in mode 2. As time progresses, population
is transferred from mode 1 to mode 2 until finally mode 1 is in
vacuum. Consequently, in phase space, mode 1 will follow an inward
spiral trajectory while mode 2 follows an outward trajectory until the
population is completely swapped between the two modes after some
time $t_s$. This is analogues to the intrinsic AHE; considering polar coordinates $r$ and $\varphi$ instead, one has that the radial motion along $r$ induces a transverse rotational current in the $\varphi$-direction. For longer time periods, the process starts over again
with the population of the two modes interchanged. The spiral motion
is depicted by the black curves in Fig.~\ref{fig2} (a) showing the
averages $\langle \hat{X}_i\rangle^2+\langle\hat{P}_i\rangle^2$ for
initial values $X_{20}=P_{10}=P_{20}=0$, $X_{10}=10$, and the atom
initially in $|1\rangle$. Figure~\ref{fig2} (b) displays the
corresponding time evolution of $\langle\hat{X}_1\rangle$ versus
$\langle\hat{X}_2\rangle$. At about 80 ns (for the utilized
parameters), the oscillations are almost entirely along the
$X_2$-axis indicating the swapping of population between the two
modes.

So far we considered a closed system neglecting system losses. From Fig.~\ref{fig2} it is clear that the two modes perform a very
large number of individual oscillations within the time $t_s$. Thus,
decoherence of both the cavity fields and the atom may become
important over such time periods. To investigate the effect of
losses, we consider the effective non-hermition Hamiltonian
$H_{eff}=H_{E\times\varepsilon}-i\kappa\left(\hat{a}^\dagger\hat{a}+\hat{b}^\dagger\hat{b}\right)-i\gamma|2\rangle\langle
2|$. Here $\kappa$ and $\gamma$ are the photon decay rate and the
atomic spontaneous emission rate respectively. The results
corresponding to the black curves of Fig.~\ref{fig2}, but where
experimental values of $\kappa$ and $\gamma$ have been considered, are shown as light blue curves
in Fig.~\ref{fig2}. The AHE clearly survives despite the relatively
long interaction time, only a slight drop in field intensity, reminiscent of momentum relaxation, is
found. It is understood that this analysis of the effect of an environment does not correctly capture decay of coherence of the system state. This, however, should not be a problem in the present situation as we are here only interested in the field amplitude of the two modes and not coherent superpositions nor the precise values of the quadrature operators.  
 
Summarizing, in this work we have shown how transverse phase space currents, reminiscent of AHEs, naturally occur in
bimodal cavity QED systems. The phenomenon has been explained in
terms of an effective non-Abelian gauge field and its corresponding
state-dependent Lorentz force acting upon the combined phase space
distributions. In particular, the effective gauge potentials induces phase space
trajectories having spiral shapes; starting in coherent states, one
field will follow an inward spiral trajectory and the other an
outward spiral trajectory causing in general a swapping of
population between the two modes. Utilizing experimental parameters,
we showed that the intrinsic AHE should be visible in current experiments even in the presence of losses.
Finally, we
point out that bimodal cavity experiments have been performed
\cite{bimodeexp}, and moreover that, apart from field intensities,
even the field quadrature operators (\ref{quad}) are easily
measured. Indeed, the ENS group of S. Haroche recently presented
experimental results were the full phase space distribution of a
cavity mode was assessed \cite{wigner}. It should also be mentioned
that the general idea presented in this work is not restricted to a
two-level atom. It equally well applies to other atom-field
configurations, such as for example a three-level $\Lambda$-atom
coupled to two cavity modes.

I wish to thank Prof. Erik Sj\"oqvist for fruitful discussions, and
acknowledge support from the MEC program (FIS2005-04627).

\end{document}